\documentclass[reprint,fleqn,superscriptaddress,twocolumn,showpacs,
amsmath,amssymb,prb,aps,short bibliography]{revtex4-1}
\usepackage[all]{xy}
\usepackage{gensymb}
\usepackage{graphicx,psfrag,times,epsfig,color}
\usepackage{verbatim,natbib}

\usepackage{color}
\usepackage{makeidx}
\usepackage{amsmath}
\usepackage{bm}
\usepackage{amsfonts}
\usepackage{amssymb}
\usepackage{hyperref}

\begin{document}

\title{The role of the pseudogap in cuprate superconductors demonstrated by the Hall effect }
\author{J\'ulia C. Anjos, H\'ercules Santana and E. V. L. de Mello}
\affiliation{Instituto de F\'{\i}sica, Universidade Federal Fluminense, 24210-346 Niter\'oi, RJ, Brazil}

\email[Corresponding author: ]{evlmello@id.uff.br}

\begin{abstract}

Cuprate high-temperature superconductors are known to have a normal-state pseudogap 
but,  after many years of intense research, its relation to the superconductivity is 
still a mystery. Similarly, the in-plane
Hall coefficient $R_{\rm H}$ has a large temperature variation
caused by changes in the carrier density which is also 
a long-standing puzzle. We approach both problem by a kinetic theory of 
phase separation that reproduces the charge density 
wave by a free energy array of potential wells.
The charge modulation favors local hole-hole attraction yielding local
superconducting pairs and a transition
by long-range phase order at $T_{\rm c}$. As the temperature increases 
the modulations decrease and the charge density increases by thermal activation.
This approach reproduces the $R_{\rm H}$ of La$_{2-x}$Sr$_x$CuO$_4$ 
compounds of $x = 0.0 - 0.25$ and clarifies the role of the pseudogap and the
disorder in cuprates.
\end{abstract}
\maketitle

The underlying physics of the pseudogap has long been probed by many different 
experiments and techniques\cite{Timusk1999} but it is still shrouded in mystery\cite{PinesPseud2005}. 
It is observed in both electron and hole-doped cuprate,
but remains unresolved despite many years of intense studies\cite{Kordyuk2015}. 
More concerning still, there is not a agreement on the 
precise phase boundaries and phenomenology of this normal state
gap. It is agreed that it represents a partial gap in the electronic
density of states\cite{Huefner2008} and is seen in 
the antinodal region of $k$-space\cite{Wise2008}. Once, it was believed
to be solely associated with the under-doped side of the phase diagram but now
it is  described as a line falling to zero either at a critical doping in the 
overdoped region\cite{Hall2016,Huefner2008} or at the end of the superconducting 
dome\cite{Huefner2008,Vishik2014}.

On the other hand, the in-plane Hall coefficient $R_{\rm H} = 1/n e$, where $e$ is the electron
charge and $n$ the itinerant carrier density, probes 
the charge dynamics of metals and semiconductors\cite{Ong1990}. 
In metals described by the Fermi liquid model, $R_H$ is constant
and negative because the conduction band density of electrons. 
In cuprate high-temperature superconductors
(HTS), it is positive due to the hole carriers and has a large variation 
with the temperature.
Comparison between powder and single crystals data demonstrated  
that the magnitude and temperature dependence of $R_{\rm H}$
is dominated by the in-plane contribution\cite{Hall.1994} but
after all these years there is not an explanation 
why $n$ varies with the temperature.

We propose here that $R_{\rm H}$ is influenced by the planar charge instabilities
that is observed in different forms like stripes\cite{Tranquada1995a}, 
checkerboard\cite{Hanaguri2004} or puddles\cite{Lang2002}. At first,
this was a phenomenon attributed solely to weakly doped compounds
around $ p = 1/8$ per Cu atom, where the charge density wave (CDW) signal is generally
more intense\cite{Ghiringhelli2012,Huecker2014,Blanco-Canosa2014}. However,
a variety of complementary experimental probes have detected charge instability 
in all hole-doped HTS families\cite{Comin2016} as well as in Nd-based electron-doped\cite{Ndoped2018}. 
More recently, CDW has been detected in overdoped La$_{2-x}$Sr$_x$CuO$_4$ 
(LSCO) up to at least $p \equiv x = 0.21$\cite{Wu2017,Shen2019,Fei2019,Tranquada2021}
and possibly up to $p = 0.27$ in the form of puddles\cite{OverJJ2022,Over2023}.
This overall presence of various forms of incommensurate charge order (CO) 
suggests that they may be more
intimately related with the superconducting phase than initially thought.

\begin{figure}[!ht] 

\centerline{\includegraphics[height=6.50cm]{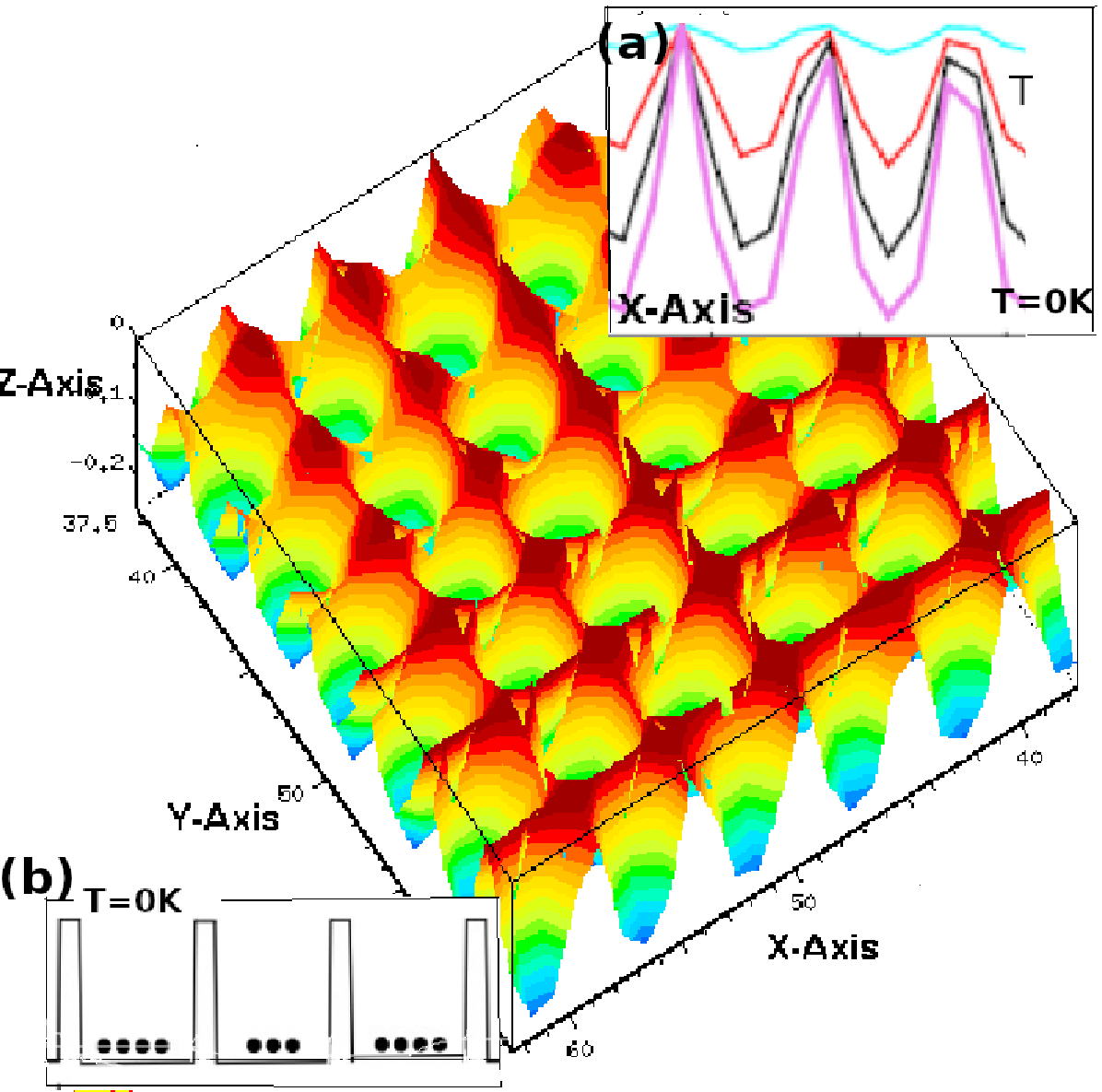}}
\caption{ Low-temperature CH simulations of the planar GL free energy phase separation 
potential with its valleys and barriers that originate the  CDW structure. Inset (a) shows the 
$V_{\rm GL}$ profile along the $x$-direction for increasing $T$ up to near $T^*$. In the 
inset (b) we describe the simplified model to the $T = 0$ K potential barriers 
to the transmission 
coefficient used to calculate the number of carriers with the temperature.
}
\label{fig1}
\end{figure}

Accordingly, we have developed a model which the pseudogap is
a mesoscopic electronic phase separation that leads to 
CO in HTS is precursor of the superconductivity\cite{DeMello2012,deMello2014,Mello2017,Mello2021,Mello2022,Mello2023}.  
We study this phenomenon by the time-dependent non-linear Cahn–Hilliard (CH)
differential equation used to describe the order-disorder transition
of binary alloys formation. The method minimizes a Ginzburg-Landau (GL) free energy given by
the (phase separation) order parameter power expansion\cite{DeMello2012,deMello2014}.
The GL potential ($V_{\rm GL}$) forms an array of wells
hosting alternate high and low hole densities domains of 
wavelength $\lambda_{\rm CO}$ size (see Fig.1).
The low density regions favor antiferromagnetic (AF) fluctuations that are 
reminiscent of the Mott AF ground state of low doping compounds and whose existence
was central to several magnetic mediated pairing theories\cite{PinesPseud2005,Huefner2008}.
When $T$ approaches $T^*(p)$, the $V_{\rm GL}$ amplitude or modulation decreases
untightening the CO, what makes, for instance, the 
softening of the x-ray peaks\cite{Huecker2014,Blanco-Canosa2014} and the
angle-resolved photoemission spectroscopy
(ARPES)\cite{Vishik2014,Kanigel2008}. We take this
effect in $R_H(p,T)$ by calculating the flow of holes tunnelling through 
the nanoscopic barriers shown in Fig. \ref{fig1} and insets.

The presence of $T^*(p)$ and the similarities between CO in hole-
and electron-doped phase diagrams\cite{Ndoped2018} suggest the existence of 
CDW modulations also around the half-filled ($p = 0$ or $n = 1$). In other words,
even the insulator compounds with $n \approx 1$ posses the $V_{\rm GL}$ or
charge modulations. Indeed, the variations of $R_H(p,T)$ at high 
temperatures\cite{Hall.2007} in compounds with very low $p$ support this
statement. Therefore, in samples with larger doping, the additional $p$ 
holes tend to fill in alternating high and low-density domains. 
Then, in a typical compound with $n = 1 + p$ itinerant carriers, the half-filled 
$n = 1$ are associated with the $V_{\rm GL}$ free energy potential
and $p$ become localized in the CDW domains in agreement with a coexisting two-gap
or two-particle scenario verified by different 
experiments\cite{Huefner2008,Wise2008,Yoshida2012,NodalGap2015}.

In this scenario, the tunnelling process through the array of $V_{\rm GL}$
and the two well-known energies of pseudogap and 
superconducting gap\cite{Huefner2008,Wise2008,Yoshida2012,NodalGap2015} are sufficient
to interpret the dynamics of the holes as a function of temperature. 
Expressing these two energies as proportional to $T^*(p)$ and the onset
of pair formation $T_c^{\rm max}(p)$,
we reproduce the main aspects of the temperature and doping dependence of $R_{\rm H}(p,T)$
of LSCO\cite{Hall.1994,Hall.2004,Hall.2006,Hall.2007}. 
These Hall effect calculations with these two aspects of the charges provide knowledge 
to understand the basic building blocks of HTS,
notably the role of the pseudogap.

The calculations start with the time-dependent CH equation which 
reproduces the CDW charge instability. It
is written in terms of the local phase separation order parameter $u (r_i,t) = (p(r_i,t) - p)/p$, where 
$p( r_i,t)$ is the local charge or hole density at a unit cell
position $i$ in the CuO plane and $p$ is the average doping level\cite{Mello2017,Mello2021,Mello2020a,Mello2020c,Mello2022}.
The simulation evolves to total phase separation, but
it is stopped when the charges reach a configuration 
close to a given CO with the $\lambda_{\rm CO}$ of a sample.
The CH equation is based on minimizing the 
GL free energy:\cite{deMelloKasal2012,deMello2014,Mello2017}: 
\begin{equation}
f(u)= {{\frac{1}{2}\varepsilon |\nabla u|^2 + V_{\rm GL}(u,T)}},
\label{FE}
\end{equation}
where $\varepsilon$ is the parameter that controls the charge modulations
$\lambda_{\rm CO}$ and ${V_{\rm GL}}(u,T)= -A^2u^2/2+B^2u^4/4+...]$ is a
double-well potential. The charge
oscillations appear below the phase separation temperature 
$T_{\rm PS}$, $B = 1$ and $A^2 = \alpha(p) [T_{\rm PS}-T]$.
In Fig. \ref{fig1} we show a typical low temperature $V_{\rm GL} (u(r_i, T))$ 
simulation and its valleys which tend to bound the carriers in alternating
high and low densities CDW domains. We show in the inset (a), the $V_{\rm GL} (u(r_i, T)$ amplitude
along the $x$-direction
that diminishes when the temperature raises towards $T_{\rm PS} \ge T^*$ 
where the system becomes homogeneous. The size of the barriers 
between the two (high and low charge density) phases is proportional to $A^4/B^2$ or 
$\langle V_{\rm GL}(p, T) \rangle = \alpha(p)^2 [T^* - T]^2$ , for $T \le T^*$. 

Thus, at low temperatures, ${V_{\rm GL}}$ produces alternating high and low charge domains and such
inhomogeneous carrier distribution induces ions` fluctuations that may lead to
hole-hole attraction\cite{Mello2020a,Mello2021}. Such indirect interaction
yields local superconducting pairs proportional to the barrier size 
$\langle V_{\rm GL}(p, T) \rangle $ what is in agreement with the HTS
coherence lengths being smaller than
the CO wavelength\cite{deMello2014,Mello2017,Mello2022}, i.e., $\xi_{\rm sc} \le \lambda_{\rm CO}$.
On the other side, when the temperature increases, $ V_{\rm GL}(p,T)$ diminishes 
as shown in inset (a) and the carriers become uniform which
leads to the temperature effect on $R_{\rm H}(p,T)$ that we will detail 
below.

We mentioned before that the $p = 0$ Hall measurements\cite{Hall.2007} also
display an increase in the number of hole carriers with the temperature similar to most
of the compounds, up to the overdoped side.
This behavior indicates that possibly all
the samples undergo a continuous phase separation below $T \approx T^*$.
At low temperatures, ${V_{\rm GL}(p,T)}$ has a series of ``ice cream cones'' as
shown in Fig. \ref{fig1} which slowly ``melts'' when the temperature raises. 
For $p \ge 0.0$, at low temperatures, the $p$ doping holes tend to become
localized or even trapped in the $n = 1$ CDW domains, where they
may also form local pairs\cite{Mello2020a,Mello2020c,Mello2022}. 
According to this scenario, we distinguish three independent 
temperature dependence to $R_{\rm H}(p,T)$; \\

First, the two sources of carriers mentioned above:\\
i- The $p$ holes of a $n = 1+ p$ compound become localized in the CDW 
alternating charge domains and, eventually, at low temperatures they form local superconducting
order parameter that establish local amplitudes and 
the superconducting state by phase coherence\cite{DeMello2012}.
The energy of these pairs are not scaled by $T_{\rm c}$ but by $T_{\rm c}^{\rm max}$, 
that is, the onset of experimental superconducting pairs or precursor pairs detected
by many experiments like ARPES\cite{Kanigel2008} and STM\cite{Gomes2007}.
Below $T_{\rm c}^{\rm max}$ the localized superconducting pairs
form an array of mesoscopic Josephson junctions, and
$R_{\rm H}$ is small. As the temperatures
increases above $T_{\rm c}$ but below $T^*$, these pairs eventually brake free
and $R_{\rm H}$ has a maximum near $T_{\rm c}^{\rm max}$. 
The number of unbound holes is obtained by thermal excitation;
$n_1(p, T) = p B_{\rm c} (1 + exp(-T_{\rm c}^{\rm max}/T))$, where
$B_{\rm c} = 2$ is the number of CuO planes 
in the unit cell of LSCO\cite{Shibauchi94}.\\

ii- The $R_{\rm H}(p,T)$ variation with the temperature\cite{Hall.2007} indicates 
that the CDW free energy modulations are already present at half-filled ($n = 1.0$ 
and $p = 0$). Therefore we assume that such modulations are formed by the half-filled holes
and not by the hole density ($p$) from doping.
Again, with the rise in temperature, 
the modulations of ${V_{\rm GL}}$ decrease as shown in the inset (a) of Fig. \ref{fig1} 
because part of the these $n = 1.0$ holes diffuse through the system.
Since the CDW sets in at $T^*$ this process is scaled by the pseudogap temperature
and the temperature dependence of the number density of such carriers is given by
thermal activation;
$n_2(p,T) = n B_{\rm c} (1 + exp(-T^*(p)/T))$, where
$n = 1$ is the {\it same} to all compounds from $p = 0$ to $p = 0.25$.\\

iii- Furthermore, all carriers have to move through
the background of the $V_{\rm GL}(p,T)$ modulations that are large at low
doping and
low temperatures but decrease when the system is warmed, as shown in 
the inset (a) of Fig. \ref{fig1}. This is taken into account  by
the tunnelling current probability between the charge domains which 
is not easy to calculate exactly, and we adopted a simplified model
based on one-dimensional quantum mechanics transmission 
coefficient\cite{QMGasiorowicz}. In this way, 
the tunnelling expression through rectangular barriers 
like those shown in the inset (b) of Fig. \ref{fig1} 
with width of $\lambda_{\rm CO}/2$ and height proportional to 
$\langle V_{\rm GL}(p, T)\rangle$ is given by.
\begin{equation}
T_{\rm trans}^2 = n(p, T) e^{ -[{\gamma\langle V_{\rm GL}\rangle (1 - (T/T^*))^2}]^{1/2}}
\label{eq2}
\end{equation}
where $n (p, T)= n_1(p, T)+n_2(p, T)$ is the total number density of holes, 
$n_1(p, T)$ and $n_2(p,T )$ are given just above,
$\gamma = 2m\hbar^2/\lambda^2_{\rm CO}$ 
and $m$ is the electron mass. The values of $ \langle V_{\rm GL}\rangle$ have been
calculated before in Refs. \onlinecite{Mello2021,Mello2022} and the ones used
in the calculations are listed
in Table \ref{table1}. Notice that $\gamma \langle V_{\rm GL}\rangle$ is adimensional
and a measurement of the CDW rougheness.

Now we use the above calculated $n (p,T)$ to reproduce the
temperature dependence of the number of holes that
is the main and most puzzling ingredient of $R_{\rm H}(p, T)$. These very
simple expressions are robust enough to describe all $ p \le 0.25$  experimental 
values\cite{Hall.1994,Hall.2004,Hall.2006,Hall.2007}
using only their values of $T^*(p)$ and $T_{\rm c}^{\rm max}(p)$. For $0.0 < p \le 0.05$ we
keep the $n (p,T)$ constant for $T \le T_{\rm AF} =300$ K, the Ne\`el temperature,
because the antiferromagnetic phase. The low density AF clusters act like as 
if the carriers are frozen what yields a plateau\cite{Hall.2007} in $R_{\rm H}(p, T)$.

The values of  $\langle V_{\rm GL}\rangle '$ $= 2m\hbar^2/\lambda^2_{\rm CO}\langle V_{\rm GL}\rangle$ = $\gamma \langle V_{\rm GL}\rangle$
for under and overdoped LSCO are obtained from the tables in the supplemental information of
Ref. \onlinecite{Mello2021} and $\lambda^2_{\rm CO}(p)$ from Ref. \onlinecite{Comin2016}.
For instance, the optimal doped $p = 0.16$ has\cite{Mello2021} $\langle V_{\rm GL}\rangle = 0. 234$ meV
and $\lambda_{\rm CO} = 3.9 a_0$ with $a_0 = 3.78 \AA $  and 
$\sqrt{\gamma \langle V_{\rm GL}\rangle} = 1.833$, and therefore we use in Eq. \ref{eq2} 
the value 1.80 to $p = 0.15$.
$\langle V_{\rm GL}\rangle '$ to other $p$'s are either calculated in this way or, 
like the small doping compounds that we do not have all the data,  are extrapolated. 
We list these results in the last column of the Table \ref{table1} below. 

\begin{figure}[!ht]
\centerline{\includegraphics[height=10.0cm]{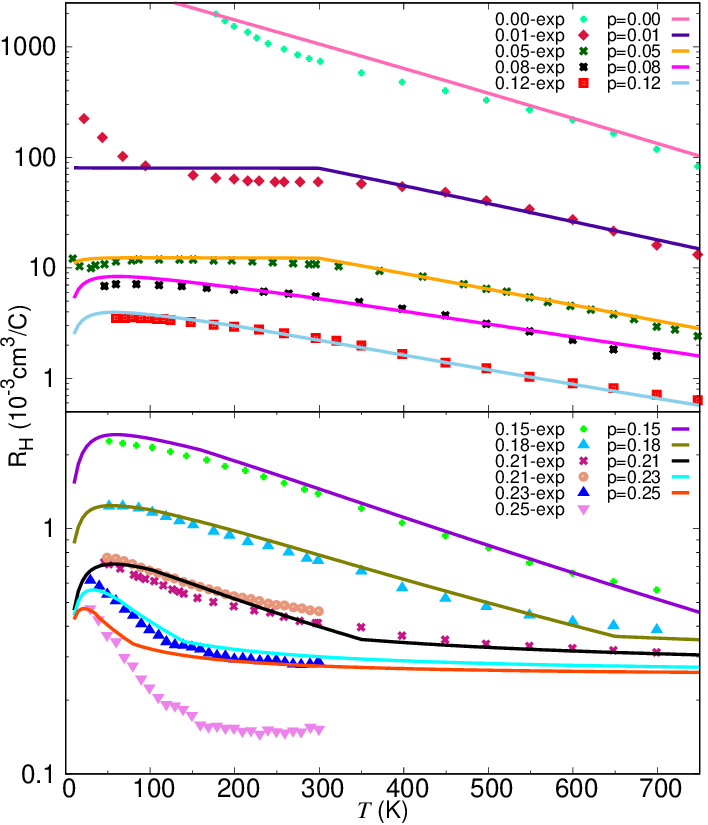}}
\caption{The $R_{\rm H}(p,T)$ 
experimental points from Ref. \onlinecite{Hall.2007} and $p =0.21$ from Ref. \onlinecite{Hall.2004} 
with the theoretical model derived in the text (the continuous lines). The agreement is good 
from $p = 0.0$ to 0.23. The case of $p = 0.25$ is the only one yielding much 
larger values above $T = 100$ K, what is most likely due to 
the negative $R_{\rm H}(p, T)$ of out-of-plane electrons in strong overdoped 
samples according to Ref. \onlinecite{Hall.2006}. 
}
\label{fig2}
\end{figure}

The $R_{\rm H}(p,T)$ results from the above analysis 
are shown in Fig. \ref{fig2} with the experimental data of
Ref. \onlinecite{Hall.2007} and the agreement is very good 
from $p = 0.0$ to 0.23. Only  $p = 0.25$ diverges from our calculations 
above $T = 100$ K. The decrease of the experimental $R_{\rm H}(p,T)$ 
is most likely due to the negative Hall coefficient of out-of planes electrons 
of strong overdoped 
samples that was reported by Ref. \onlinecite{Hall.2006}. The presence of
electrons may be connected with the decrease of the rougheness parameter below 1.0
shown in Table \ref{table1}.

\begin{table}[!ht]
\caption{ Properties of LSCO used in the calculations. The first column is 
hole doping density per unit cell. 
Second is the $T^*$ and third is $T_{\rm c}$ that does not enter in the 
calculations but is a good reference and the last column is the renormalized free
energy barrier $\sqrt( \langle V_{\rm GL}\rangle')$ or rougheness factor explained
in the text.
}
\begin{tabular}{|c|c|c|c|c|}\hline \hline
Sample &  $T^*$ (K) & $T_{\rm c}$ (K) & $T_{\rm c}^{\rm max}$ (K) 
& $\sqrt(\langle V_{\rm GL} \rangle')$   \\ \hline
 {\it p} = 0.00 & {\bf 1800} &  0.0 & {\bf 0.0 }& {\bf 9.0 }   \\ \hline
 {\it p} = 0.01 & {\bf 1700} & 0.0 & {\bf 10.0  }& {\bf 6.0} \\ \hline
 {\it p} = 0.05 & {\bf 1350} & 1.0  &{\bf 70.0 }& {\bf 4.0} \\ \hline
 {\it p} = 0.08 & {\bf 1300}  & 20.0 &{\bf 280.0} &{\bf 3.0} \\ \hline
 {\it p} = 0.12 & {\bf 900} & 34.0 &{\bf 200.0 }&  {\bf 2.3} \\ \hline
 {\it p} = 0.15 & {\bf 850} & 39.0  &{\bf 160.0  }& {\bf 1.8}\\ \hline
 {\it p} = 0.18 & {\bf 650} & 36.0  & {\bf 160.0 } & {\bf 1.1}\\ \hline
 {\it p} = 0.21 & {\bf 350} & 24.0  &{\bf  110.0  }& {\bf 0.6}\\ \hline
 {\it p} = 0.23 & {\bf 140} & 18.0  &{\bf  70.0 } & {\bf 0.4}\\ \hline
 {\it p} = 0.25 & {\bf 80} & 12.0  & {\bf 40.0 } & {\bf 0.2}\\ \hline
 \end{tabular}
 \label{table1}
 \end{table}

In conclusion, the pseudogap`s role in cuprate superconductors as a 
mediator of phase separation
and the two-particle and two-gap scenario\cite{Huefner2008} are identified by the
$R_{\rm H}(p, T)$ calculations. 
The model based on phase-ordering kinetics\cite{Mello2021} reveal the nature of the two types of 
carriers in a typical compound of $n = 1 + p$ hole doping: The
half-filled charges ($n = 1$) of a single band are responsible by 
the phase separation free energy shown in Fig. \ref{fig1} and
scaled by the temperature $T^*(p)$. The other $p$ holes of a $n = 1 + p$ compound
fill in the alternating CDW domains breaking the uniform charge
distribution and inducing ion-mediated pair interaction. The energy of
these pairs are scaled by the onset temperature
of superconducting pair formation\cite{Gomes2007,Kanigel2008} $T_{\rm c}^{\rm max}(p)$,
that is just a few times $T_{\rm c}$ (see Table \ref{table1}). 
For low doping the physical properties like superfluid density or quantum 
oscillations\cite{Hall2016,Mello2020c} are dominated by the thermal activated $n = p$ holes
because $T_{\rm c}^{\rm max}(p) \ll T^*(p)$.
Increasing doping, above $p \approx 0.19$
the two energies or temperatures converge to the 
same order of magnitude\cite{Shen2019} (see Table \ref{table1}) 
and the carrier concentration detected in experiments crosses over from $p$ to
$1 + p$\cite{Hall2016,OverJJ2022,Over2023,Mello2020c}.
Taking into account the above temperature dependent contributions we reproduced {\it all}
the LSCO $R_{\rm H}(p,T)$ measurements.

We acknowledge partial support from the Brazilian agencies CNPq and 
by Funda\c{c}\~ao Carlos Chagas Filho de Amparo
Pesquisa do Estado do Rio de Janeiro (FAPERJ), Projects No.
E-26/010.001497/2019 and No. E-26/211.270/2021.

%

\end{document}